\documentstyle[aps,epsfig,prl,floats]{revtex} 
\begin{document}
\draft
\twocolumn[\hsize\textwidth\columnwidth\hsize\csname@twocolumnfalse%
\endcsname
\title{Spin and Current Correlation Functions in the
$\bf d$-density Wave State of the Cuprates}
\author{Sumanta Tewari, Hae-Young Kee, Chetan Nayak, and Sudip Chakravarty}
\address{Physics Department, University of California, Los Angeles, CA 
  90095--1547}
\date{\today}
\maketitle
\begin{abstract}
We calculate the spin-spin and current-current correlation
functions in states exhibiting $d_{{x^2}-{y^2}}$-density wave
(DDW) order, $d_{{x^2}-{y^2}}$ superconducting order (DSC),
or both types of order. The spin-spin correlation functions in a state
with both DDW and DSC order and in a state 
with DDW order alone, respectively, illuminate the resonant peak
seen in the superconducting state of
the underdoped cuprates and the corresponding
feature seen in the pseudogap regime.
The current-current
correlation function in a state
with both DDW and DSC order evinces a superfluid density
with doping dependence which is consistent with that
of the underdoped cuprates. These calculations
strengthen the identification of the pseudogap
with DDW order and of the underdoped cuprates
with a state with both DDW and DSC order.
\end{abstract}

\pacs{PACS numbers: }
]
\narrowtext
%\newpage
\section{Introduction}
Recently, a new order termed $d$-density wave 
(DDW) was proposed for the
underdoped regime of the high-$T_{c}$ cuprate
superconductors\cite{Sudip,Chetan,CK}. It was argued 
that the observed anomalies in the 
pseudogap gap phase are due to the DDW and its competition
with the $d$-wave superconductor
(DSC). Recent {\it elastic} neutron scattering \cite{Mook-DDW},
$\mu$SR \cite{Sonier}, and NMR/NQR\cite{Dooglav} experiments
appear to have found some {\it direct}
evidence for the existence of DDW order in
YBCO by observing the magnetic fields
associated with this order parameter -- in other
words, by measuring the order parameter itself.
In this paper, we wish to elucidate the nature
of the competition between the DDW and DSC
order paramters by explicitly
computing the spin-spin and the current-current
correlation functions in two space-dimensions in the
Hartree-Fock approximation. Our aim is to provide
complementary, {\it indirect} evidence in favor
of the DDW scenario by showing in some detail that
it provides a consistent explanation of the existence
and behavior of the resonant peak seen in {\it inelastic}
neutron scattering and also of the evolution of the
superfluid density with doping and temperature, at low
temperatures.

Since we deal with
broken symmetry states with order
parameters, it is possible to argue that their
descriptions are adequately described by the
Hartree-Fock approximation, at least deep within
the ordered phases at zero temperature.
The Hartree-Fock approximation by
itself cannot be adequate if
collective modes are important, but these modes could be described
in terms of small oscillations about the
ordered state by suitable random phase approximations
(RPA). However, critical fluctuations close to
a quantum critical point cannot be treated by  either Hartree-Fock or RPA.
Nonetheless, it is useful to explore the
consequences of the simplest possible Hartree-Fock and RPA
theories to see if any robust features  arise
due to the existence of the new order parameter, DDW,
and its competition with DSC.

For underdoped bilayer superconductors \cite{FongMook},
the dynamic spin-spin correlation function
is peaked above $T_{c}$ at the in-plane wavevector
${\bf Q}=(\pi,\pi)$ (in this paper we shall take the lattice
constant to be unity and will also set
$\hbar = c = k_{B} = 1$) in both even and odd
channels with respect to the layers in the bilayer complex.
Below $T_c$, there is a resonant peak at the same wavevector
and approximately the same energy in the odd channel. A resonant
peak is also found below $T_c$ in optimal and overdoped bilayer
high-$T_c$ superconductors.
There are many
approximate theoretical calculations of the dynamic structure
factor\cite{resonance}. Here we concentrate only on the aspects germane to our
discussion of DDW. We wish to stress that this analysis is simply a consistency
check of the DDW; we do not claim that
our explanation of the resonant peak in inelastic neutron scattering
is unique or better than the others. The
real test of the DDW hypothesis
is the direct observation of the order parameter referred
to earlier\cite{Mook-DDW,Sonier,Dooglav}. We find that the
experimental observations of the neutron dynamic structure
factor are consistent with the hypothesis that the neutron
scattering peak -- and, hence, the pseudogap -- is due to
DDW order, while the resonance peak in the underdoped
superconducting state is due to the coexistence of DDW and
DSC orders\cite{CK}.  The suppression of the resonant peak
by a magnetic field perpendicular to the plane suggests that
superconducting pairing is important for its
formation\cite{Mook}.  

The behavior of the superfluid density in the
underdoped regime of the high-$T_{c}$
materials is another interesting physical quantity.
Many experiments\cite{Loram}
have indicated a rapid collapse 
of the zero-temperature superfluid density as the doping
is decreased below optimal. The zero-temperature
penetration depth, for example,
grows rapidly in the pseudogap regime, correlating
with the suppression of $T_{c}$
with underdoping, yet saturates at overdoping in
a way reminiscent of a traditional 
BCS superconductor\cite{Uemura}. 
This, as we will show, can be explained by the competition
between DDW and DSC orders.
At finite temperatures the superfluid density is
suppressed from its zero temperature value at that doping fraction. This
suppression is  linear in temperature ($T$) at asymptotically low temperatures,
with a slope that is independent of doping. This behavior is  captured in our
calculations by the thermal excitation of nodal quasiparticles into the upper
quasiparticle band. For heavily underdoped samples, there is an intermediate low
temperature regime in which we predict this suppression to be proportional to
$\sqrt{T}$. There is some experimental evidence for this behavior\cite{Panago}in
YBa$_2$Cu$_4$O$_8$, although the explanation of these authors involves  a proximity
model of alternating stacked superconducting and normal layers. Our analysis is 
simpler and follows  from the nodal exciations in the mixed DDW and DSC state. This
prediction can be tested in future  experiments in the regime in which pseudogap
dominates the DSC gap.

\section{Order Parameter}
Our Hartree-Fock analysis of the physical quantities merely requires us to
specify a mean field Hamiltonian with the proposed order parameters. The actual
microscopic Hartree-Fock analyses to obtain these order parameters have been
discussed in the past on numerous occasions and need not be repeated here;
for a recent set of calculations, see Refs.~\cite{Chetan},
\cite{Zeyher}, and references therein.  
Hamiltonians with short range repulsion and superexchange have
the DDW ordered state (both the singlet and the triplet variety) as one of
their many possible saddle points.
If one includes correlated hopping terms,
then within a reasonable range of parameters,
the DDW saddle point can be stabilized 
over the other possibilities.
In the present paper, we
shall parametrize the order parameters phenomenologically, and our main
conclusions are independent of any microscopic self-consistent Hartree-Fock
equations.
     
The singlet DDW state is defined by the  order parameter\cite{Chetan}
\begin{equation}
{\langle c_{{\bf k} \alpha}^{\dagger}}
{c^{}_{{\bf k+q} \beta}}\rangle=
i\frac{\Phi_{\bf q}}{2}(\cos k_{x}-\cos k_{y}) \delta_{\alpha\beta},
\end {equation}
where $c_{ {\bf k} \alpha}$ is the fermion destruction operator for wavevector
${\bf k}$ and spin-index $\alpha$,
and ${\bf q}$ is the ordering wavevector.
This order parameter is a
generalization of the familiar charge density wave 
(CDW) order parameter to the case of angular momentum two, in
exact analogy to the generalization of the BCS $s$-wave superconducting order 
parameter to its $d$-wave (DSC) counterpart. 

For ${\bf q}={\bf Q}$, the case
of interest to us, the underlying bipartite square lattice band structure
is equivalent under the transformation ${\bf Q}\rightarrow -{\bf Q}$ and this
forces the DDW order parameter to be imaginary \cite{Chetan}.
Thus, time-reversal symmetry is broken
({\it i.e.} the system exhibits magnetism), and the ground state
has an array of bond currents, which alternate
in direction (clockwise, anticlockwise) in
the neighboring plaquettes of the 2D
bipartite square lattice. The corresponding $s$-wave
CDW order parameter is, of course, real.

\section{Spin Dynamics}
\subsection{Pure DDW state}
  
The Hartree-Fock DDW Hamiltonian, using (1), is
\begin{equation}
H_{\rm DDW} = \sum_{{\bf k} \sigma}
( \epsilon_{\bf k} - \mu ) c^{\dagger}_{  {\bf k} \sigma}
c_{  {\bf k} \sigma} + \sum_{{\bf k} \sigma} 
i W_{\bf k} c^{\dagger}_{  {\bf k } \sigma}
c_{  {\bf k +Q} \sigma} + \mbox{h.c.} , 
\end{equation}
where the DDW gap is given by
\begin{equation}
W_{\bf k}=\frac{W_0}{2}(\cos k_{x}-\cos k_{y}) ,
\end{equation}          
$\epsilon_{ {\bf k}}=-2t(\cos k_{x}+\cos k_{y})$
gives the band structure, and $\mu$ is the
chemical potential. A more realistic
band structure would include
the effect of next-neighbor hopping $t'$.
This does not affect our results here, so
we drop it for simplicity. At half filling, the
chemical potential $\mu=0$, while $\mu$ takes
non-zero negative values  as we introduce holes into the system. 
 
At $\mu=0$ the zero-temperature spin-spin
correlation function $\cal S$,
at momentum transfer ${\bf q} ={\bf Q}$, is given by
 \begin{equation}
 {\cal S}({\bf Q},\omega,\mu=0)=3\pi\sum_{{\bf k}\in {\rm rbz}}
\delta\left(\omega-2\sqrt{W_{{\bf k}}^{2}+
 \epsilon_{{\bf k}}^{2}}\right),
\label{eq:smu0}
 \end{equation}
where the integration is over the magnetic or reduced Brillouin zone (rbz).
This function is peaked at an energy equal to twice
the maximum value of the gap, $2W_0$.
As we dope holes into the system, at nonzero
values of $\mu$, Eq.~\ref{eq:smu0} is changed to
\begin{eqnarray}
{\cal S}({\bf Q},\omega,\mu)=
3\pi\sum_{{\bf k}\in {\rm rbz}}&\delta&\left(\omega-2\sqrt{W_{{\bf
k}}^{2}+\epsilon_{{\bf k}}^{2}}\right)\nonumber \\
&\times&\theta\left(\mu+\sqrt{W_{{\bf k}}^{2}+\epsilon_{{\bf k}}^{2}}\right)
\label{eq:sddw}
\end{eqnarray}
The only effect of $\mu$ is the depletion of
the integration region. Hence, as long
as $|\mu|<W_0$, the peak at ${\bf q}={\bf Q}$ exists
and stays at the same energy $2W_0$.

Note that ${\cal S}(\omega)$ also shows a peak for momentum transfer
${\bf q}=(\pi,0)$ \cite{Affleck} at the Hartree-Fock level.
{}From a relation analogous to that of
Eq.~\ref{eq:sddw}, we can see that
the peak-energy for ${\bf q}=(\pi,0)$ scales with $t$.
This is much higher in energy than the $(\pi,\pi)$
peak; it is likely to have an enormous width
(since there is an enormous amount of phase space into
which it can decay) and be unobservable. This
is important because no peak has been observed
at ${\bf q}=(\pi,0)$ in the spin-fluctuation spectrum
of the pseudogap regime of the high-$T_{c}$ materials.
Early mean-field decouplings of the Heisenberg model
\cite{Affleck} led to an effective $t \sim J$, so that the
peaks at ${\bf q}=(\pi,\pi)$ and ${\bf q}=(\pi,0)$
were comparable in energy, in contradiction to the
experiments.

\subsection{Pure DSC state}
 
Below the superconducting transition
temperature $T_{c}$, a DSC state is defined 
by the gap 
 \begin{equation}
  \Delta_{ {\bf k}}=\frac{\Delta_0}{2}(\cos k_{x}-\cos k_{y})
 \end {equation}
corresponding to the DSC order parameter.
This order parameter, as stressed in reference\cite{Sudip}, can compete and
coexist  with the singlet DDW order parameter.
In fact, at half filling, the system can be rotated continuously
from a pure DSC order to a pure singlet DDW order
and {\it vice versa}, without 
ever having to close the quasi-particle gap \cite{Sudip}. (The
symmetry between the $s$-wave counterparts of these two types of
orders is exact in the negative $U$ Hubbard model \cite{Fradkin}.)

Using (6), the mean-field one-body DSC Hamiltonian is
\begin{equation}
H_{\rm DSC} = \sum_{{\bf k} \sigma} ( \epsilon_{\bf k} - \mu )
c^{\dagger}_{  {\bf k} \sigma}
c_{  {\bf k} \sigma} + \sum_{{\bf k}}
\Delta_{{\bf k}}c_{{\bf k}\uparrow}^{\dagger}
c_{{-\bf k}\downarrow}^{\dagger} + \mbox{h.c.}
\end{equation}
The spin-spin correlation function,
at $\mu=0$ and momentum transfer ${\bf q}$, is
\begin{eqnarray}
   {\cal S}({\bf q},\omega,\mu=0)&=& \frac{3\pi}{2}
\sum_{{\bf k}\in {\rm rbz}}\left(1-\frac{\epsilon_{{\bf
k}}\epsilon_{{\bf k+q}}+
   \Delta_{{\bf k}}\Delta_{{\bf k+q}}}{E_{{\bf k}}
E_{{\bf k+q}}}\right) \nonumber\\
   &\times& \delta(\omega-E_{{\bf k}}-E_{{\bf k+q}}),
\label{eq:sdsc}
 \end{eqnarray}
 where $E_{{\bf k}}=\sqrt{\Delta_{{\bf k}}^{2}+
\epsilon_{{\bf k}}^{2}}$. For ${\bf q}={\bf Q}$
 the coherence factor equals two and
Eq.~\ref{eq:sdsc} becomes identical to Eq.~\ref{eq:smu0}.
Hence, just like the DDW, at $\mu=0$ 
the DSC spin-spin correlator also shows a peak at 
twice the gap maximum, $2\Delta_{0}$.
  
But for a DSC, at non-zero values of $\mu$,
${\cal S}({\bf q},\omega,\mu)$ is given by
\begin{eqnarray}
 {\cal S}({\bf q},\omega,\mu)&=& \frac{3\pi}{2}\sum_{{\bf k}\in {\rm rbz}}
\left(1-\frac{\widetilde{\epsilon}_{\bf k}\widetilde{\epsilon}_{\bf k+q} 
   +\Delta_{\bf k}\Delta_{\bf k+q}}{E_{\bf k}E_{\bf k+q}}\right)\nonumber \\ 
   &\times&\delta(\omega-E_{{\bf k}}-E_{{\bf k+q}}) ,
\label{eq:sdscmu}
\end{eqnarray}  
where $E_{\bf k}$ now incorporates $\mu$
in the usual way by $\epsilon_{{\bf k}}
\to
\widetilde{\epsilon}_{\bf k} = \epsilon_{\bf k}-\mu$.
Equation ~\ref{eq:sdscmu} shows a peak at ${\bf q}={\bf Q}$ that is shifted 
in energy from $2\Delta_{0}$. The peak-energy shift
is small for small $|\mu|$ ,
but can be easily checked to scale with $2|\mu|$ as $|\mu|$ becomes large.
Hence, one can clearly see that the peak-energy behaves differently with $\mu$ 
for DDW and DSC. 

These Hartree-Fock calculations for the resonance peak in the normal state becom
e interesting
only if the chemical potential remains pinned close to $\mu=0$ in
the underdoped regime.
(This is not true for the calculation of the superfluid density reported in
the following section, which is insensitive to such assumptions for $\mu$.)
There is some evidence that
this is the case in the photoemission measurements of  Ino {\it et al.}\cite{Ino} in
$\mbox{La}_{2-x}\mbox{Sr}_{x}\mbox{CuO}_{4}$.
They find that while the chemical potential shift is large
in the overdoped samples, it is largely suppressed in the
underdoped regime. Numerical studies of the
2D Hubbard model\cite{Furukawa,Dagotto} also suggest
that the chemical potential
does not shift much for small doping-fractions.
In the Monte Carlo study
performed in reference \cite {Furukawa}, the calculated
shift of $\mu$ follows $\Delta\mu = \mu \propto - x^2$ for
small values of doping, $x$.
The data of reference\cite{Ino} suggest that the chemical potential varies
with $x$ in the overdoped regime as one would expect for a Fermi
liquid, but not in the underdoped regime.
Ideas involving charge ordered stripe states\cite{Steve}
are suggestive of the segregation of doped holes in the boundaries
of antiferromagnetic domains, thus pinning the
chemical potential. The presence of charge inhomogeneity,
such as stripes\cite{Kivelson}, or impurity-disorder
may pin the
chemical potential in other
multilayer high-$T_{c}$ cuprate systems as well.

\subsection{Coexisting DDW and DSC}     
 \begin{figure}[bht] 
   \narrowtext 
   \begin{center} 
   \leavevmode 
%\vspace{.2cm} 
   \noindent 
   \hspace{0.3 in} 
   \centerline{\epsfxsize=2.7in \epsffile{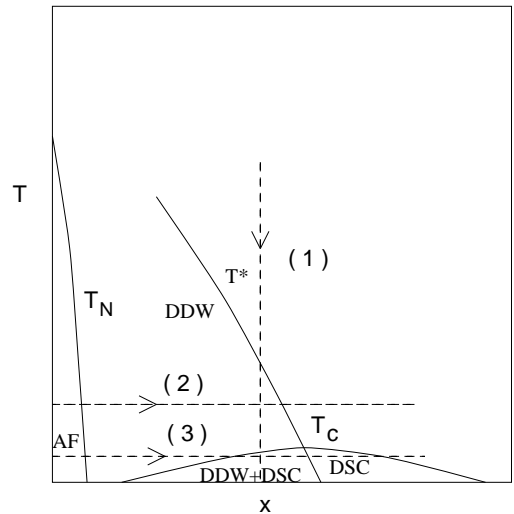}} 
%\epsfxsize=3.3in 
%\epsfysize=2.in 
%\epsfbox{} 
%\vspace{.2cm} 
   \end{center} 
   \caption 
{The high-$T_{c}$ cuprate phase diagram proposed in Ref. 1. $T_N$ denotes the 
3D-
antiferromagnetic transition temperature. $T_{c}$ and $T^{*}$ are the DSC and 
pseudogap, or DDW,
transition temperatures, respectively.   In the text (Sec. V), the behavior of
the spin-s pin 
correlator
is discussed along the three paths shown here by the dashed lines.} 
   \label{fig:3} 
   \end{figure}
    
{}From the order parameter competition picture proposed
in reference\cite{Sudip} (See, Fig.~\ref{fig:3}), one notes that
to the right of $x\simeq 0.05$ and below $T_{c}$, the two orders,
singlet DDW and DSC, coexist 
and compete for the same regions on the Fermi
surface. The mean-field Hamiltonian for the system in the
folded Nambu basis\cite{Schrieffer} is
\begin{equation}
H-\mu N = \sum_{{\bf k}\in {\rm rbz}}
\Psi_{\bf k}^{\dagger} A_{\bf k}\Psi_{\bf k
},
\label{eq:ddwdsc}
\end{equation}
where $\Psi_{\bf k}^{\dagger}= (c_{{\bf k}\uparrow}^{\dagger}, 
c_{{\bf k+Q}\uparrow}^{\dagger}, 
c_{{\bf -k}\downarrow}, c_{{\bf -k-Q}\downarrow})$ and the matrix 
$A_{\bf k}$ is

\begin{equation}
   A_{\bf k}=\left( \begin{array}{cccc} (\epsilon_{\bf k}-\mu) & iW_{\bf k} & 
    \Delta_{{\bf k}} & 0 \\
  -iW_{{\bf k}} & -(\epsilon_{\bf k}+\mu) & 0 & -\Delta_{\bf k} \\
  \Delta_{\bf k} & 0 & -(\epsilon_{\bf k}-\mu) & iW_{{\bf k}} \\
   0 &-\Delta_{\bf k}  & -iW_{\bf k} & (\epsilon_{\bf k}+\mu) \\
    \end{array} \right) 
\end{equation}
The matrix $A_{\bf k}$  has four eigenvalues,
$\pm E_{1}({\bf k})$ and $\pm E_{2}({\bf k})$, where
$ E_{1}({\bf k})=[(\sqrt{\epsilon_{{\bf k}}^{2}+W_{{\bf k}}^{2}}-\mu)^{2}+
\Delta_{{\bf k}}^{2}]^{1/2}$
and $E_{2}({\bf k})=
[(\sqrt{\epsilon_{{\bf k}}^{2}+W_{{\bf k}}^{2}}+\mu)^{2}
+\Delta_{{\bf k}}^{2}]^{1/2}$. 

The spin-spin correlation function is obtained by first
evaluating the imaginary-time ordered correlator ($\omega_m$ is the bosonic 
Matsubara frequency)
\begin{eqnarray}
\nonumber\langle T_{\tau}&&{\bf S}({\bf q},
i\omega_{m})\cdot{\bf S}(-{\bf q}, -i\omega_{m})\rangle=
    \\ && \frac{3}{4\beta}\sum_{n,{\bf k}\in {\rm rbz}}  {\rm Tr\ }
\left[G({\bf k}, i\omega_{n})
 G({\bf k} + {\bf q}, i\omega_{n} + i\omega_{m}) \right]
\end{eqnarray}
where ${\bf S}$ denotes the spin operator and the $G$'s are the 4$\times$4 
matrix Green functions computed from the Hamiltonian in Eq.~\ref{eq:ddwdsc}. 
After analytically
continuing the frequency to the real axis, we extract the imaginary part
and use the fluctuation-dissipation theorem to get the spin-spin correlation
function. 

The zero-temperature spin-spin correlation function thus obtained,
at non-zero $\mu$, for momentum transfer ${\bf q}={\bf Q}$, is given by
\begin{eqnarray}  
{\cal S}({\bf Q},\omega,\mu)= \frac{3\pi}{2}
\sum_{{\bf k}\in {\rm rbz}}&&\left(1+\frac{\epsilon_{{\bf k}}^{2}+
\Delta_{{\bf k}}^{2}+W_{{\bf k}}^{2} 
-\mu^{2}}{E_{1}({\bf k})E_{2}({\bf k})}\right)\nonumber \\
   &\times&\delta\left(\omega-E_{1}({\bf k})-E_{2}({\bf k})\right)
\label{eq:sdscddw}
\end{eqnarray}
When $W_{{\bf k}} = 0$, the coherence factor
in Eq.~\ref{eq:sdscddw} clearly matches with
the coherence factor in Eq.~\ref{eq:sdscmu},
where we have to put ${\bf q}={\bf 
Q}$, and the two
expressions become identical. On the other hand, for $\Delta_{{\bf k}} = 0$,
one notes that $ E_{1}({\bf k})\to |\sqrt{\epsilon_{{\bf k}}^{2}
+W_{{\bf k}}^{2}}-\mu |$, and $ E_{2}({\bf k})\to |\sqrt{\epsilon_{{\bf k}}^{2}
+W_{{\bf k}}^{2}}+\mu |$. Keeping in mind that $\mu$ is negative in our 
hole-doped system, we note that for 
$\sqrt{\epsilon_{{\bf k}}^{2}+W_{{\bf k}}^{2}}+\mu > 0$ 
the coherence factor in Eq.~\ref{eq:sdscddw} is two, while
the other choice
for $\sqrt{\epsilon_{{\bf k}}^{2}+W_{{\bf k}}^{2}}+\mu < 0$  reduces the 
coherence factor to zero. Thus the expression in
Eq.~\ref{eq:sdscddw} becomes identical to the expression
in Eq.~\ref{eq:sddw} for $\Delta_{{\bf k}} = 0$.

One can easily check that ${\cal S}({\bf Q},\omega,\mu)$, as a function of 
$\omega$,
has a peak in the spin-spin correlation function . 
The peak is
located at $2\sqrt{W_{0}^{2} + \Delta_{0}^{2}}$ for $\mu = 0$. 
For small $\mu$ the peak-energy is shifted to higher values, 
initially by a small amount, but finally
scaling as $2|\mu|$ for large values of $|\mu|$. We should
mention here that one can in fact control
the behavior of the peak-energy with $\mu$ by adjusting the relative 
strengths of
the order parameters. When the singlet DDW order overshadows the DSC order, 
the peak-energy will tend to be pinned to the total order parameter 
magnitude $2\sqrt {W_{0}^{2} + \Delta_{0}^{2}}$, 
a behavior characteristic of the singlet DDW.

\subsection{RPA Corrections}

\begin{figure}[bht] 
\narrowtext 
\begin{center} 
\leavevmode 
%\vspace{.2cm} 
\noindent 
\hspace{0.3 in} 
\centerline{\epsfxsize=2.7in \epsffile{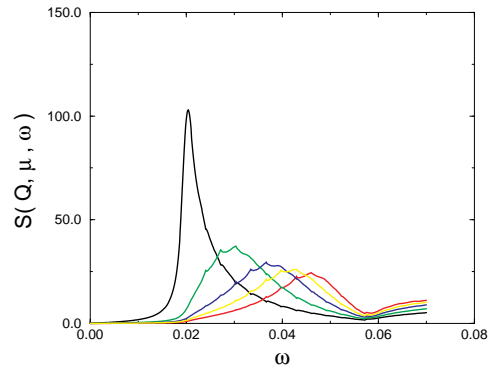}} 
%\epsfxsize=3.3in 
%\epsfysize=2.in 
%\epsfbox{} 
%\vspace{.2cm} 
\end{center}
 
\caption 
{${\cal S}({\bf Q},\omega,\mu)$ as a function
of $\omega$ after $\overline U$ is introduced into the system with 
coexisting singlet DDW and DSC orders with $t=0.5$ eV,
$\Delta_{0}$=.02 eV, $W_{0}$=.02 eV
and $\mu$ set to $-.01$ eV. As $\overline U$ is increased,
the structures are peaked at lower 
and lower energies; from right to left ${\overline U}=0.2, 0.217,
0.233, 0.256, 0.284$. Finally they evolve, as ${\overline
U}={\overline U}_{c}=0.32 eV$ is approached, into a divergence at
zero energy.} 
\label{fig:1} 
\end{figure}

If we add a reduced Coulomb repulsion $\overline U$,
which is assumed to be renormalized
due to particle-particle 
correlations, we can
drive the system toward antiferromagnetism. 
To model this effect, we have to go beyond the Hartree-Fock, and  we use a crude
RPA form\cite{Sudip} for the susceptibility
$\chi({\bf q},\omega)$ given by,
\begin{equation}
\chi({\bf q},\omega) =
\frac {\chi_{0}({\bf q},\omega)}{1- {\overline U}\chi_{0}({\bf q},\omega)}
\end{equation}
to describe the spin dynamics in the system.
Here $\chi_{0}({\bf q},\omega)$ is the
Hartree-Fock susceptibility. Extracting ${\cal S}({\bf q},\omega,\mu)$
from $\chi({\bf q},\omega,\mu)$,
in Fig.~\ref{fig:1} we plot the results as a function
of frequency for five different values of $\overline U$.
 
The first thing one notices is that after the introduction of $\overline U$,
the correlation function ${\cal S}$, which was
logarithmically singular when derived
from $\chi_{0}({\bf q},\omega)$ alone, now becomes broad. Also, as 
$\overline U$ is progressively increased, the spectral-weight appears 
at smaller and smaller energies. For $\overline U$ large enough,
the correlator starts 
peaking up again and finally evolves into a divergence at zero energy
as ${\overline U}_{c}$ is approached. All of this behavior is qualitatively
consistent with what happens to the spin-fluctuation 
spectrum as the doping is reduced in the underdoped regime of the
cuprates. As the  hole density is reduced, the repulsive Coulomb energy 
becomes more and more
important and gradually approaches the critical
value ${\overline U}_{c}$ for which the peak 
in the spin-spin correlation function evolves into a divergence 
at zero energy, signalling
antiferromagnetic order in the system.

This analysis applies both below and above $T_{c}$ in the high-$T_{c}$
cuprate materials. Above the superconducting transition temperature
$T_{c}$, $\Delta_{{\bf k}}$ will go to
zero, but $W_{{\bf k}}$ will survive up to the pseudogap temperature
$T^{*}$. Hence, in the spin-spin correlator
the DDW will continue to maintain its signature intact in the form of a peak 
at twice the maximum of its gap.

\section{Superfluid Density}

\begin{figure}[bht] 
   \narrowtext 
   \begin{center} 
   \leavevmode 
%\vspace{.2cm} 
   \noindent 
   \hspace{0.3 in} 
   \centerline{\epsfxsize=2.7in \epsffile{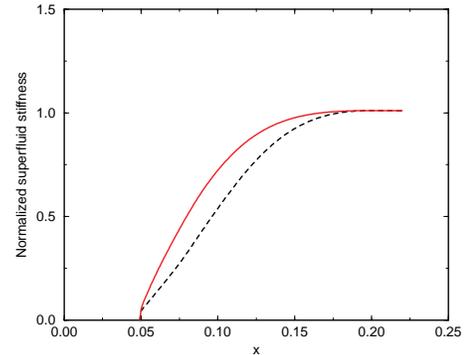}} 
%\epsfxsize=3.3in 
%\epsfysize=2.in 
%\epsfbox{} 
%\vspace{.2cm} 
   \end{center} 
   \caption 
{The $T=0$ superfluid stiffness $n_s/m^*$ as a function of doping fraction $x$. The 
vertical axis is normalized by the superfluid stiffness at $x=0.2$. Here, 
we have set $t=0.5$ eV and the chemical potential is taken to be
$\mu =-x^2$. The  amplitudes of the order parameters
are as given in Eqs.~\ref{eq:doping} and
\ref{eq:W0}. The dashed curve is $n_s/m^*$ for 
$W_{0}$ dispersing quadratically with $x$ as explained in the text.} 
   \label{fig:2} 
   \end{figure}

The anomalous behavior of the superfluid density in the underdoped regime of 
the high-$T_{c}$ cuprates has been discussed extensively. Here, we address 
it in the context of the competition of the order parameters 
that we have been discussing\cite{Dunghai}.
We will make a phenomenological assumption about
the doping dependence of the DDW and single particle gaps,
respectively, $W_{{\bf k}}$ and $\sqrt{\Delta_{{\bf k}}^2 + W_{{\bf k}}^2}$.
Following \cite{Loram2}, we hypothesize that the DDW
order parameter vanishes above a critical
doping $x_c$ (about 0.2) and follows the
formula $W_{0}/k_B\simeq\alpha J^{*}(1-x/x_{c})$
where $J^{*}=980 K$, $x$ is the doping-fraction,
and $\alpha\simeq0.6$ is a constant arbitrarily
chosen to set a scale. In other words,
\begin{equation}
  W_{0}\simeq.049(1-x/x_{c}) 
\label{eq:doping}
\end{equation}
in eV; in Fig.~\ref{fig:2}, we test another model in which $W_0\propto (1-x/x_c)
^2$.

We also posit that the single-particle gap remains constant at its value at 
$x=.05$, where superconductivity starts developing; then
\begin{equation}
  W_{0}^{2}+\Delta_{0}^{2} \equiv \Phi_0^2 =(.037)^{2}
\label{eq:W0}
\end{equation}

We calculate the superfluid stiffness for such a model
using the formula\cite{Scalapino}
\begin{equation}
\frac{n_s}{m^*}=\langle-k_{x}\rangle-\Lambda_{xx}(q_{x}=0,q_{y}
\rightarrow 0,i\omega_{m}=0)
\end{equation}
which gives the response to a transverse
vector potential $A_{x}(t)$. Here,
$n_s$ is the 
superfluid density and $m^*$ is the effective mass, $\langle k_{x}\rangle$
is the kinetic energy per site per lattice dimension,
and $\Lambda_{xx}({\bf q},i\omega_{m})$ is the paramagnetic current-current
correlation function. 

The paramagnetic current response
is obtained from the relation
\begin{eqnarray}
  \nonumber\Lambda_{xx}({\bf q}, i\omega_{m})&=&
   \langle T_{\tau}j_{x}({\bf q}, i\omega_{m})j_{x}(-{\bf q}, -i\omega_{m})
    \rangle\\&=&
    \nonumber \frac{4t^{2}}{\beta}
   \sum_{n,{\bf k}\in {\rm rbz}}\sin k_{x} \sin ({\bf k+q})_{x}\\&& 
   \nonumber{\rm Tr}[G({\bf k}, i\omega_{n})M
    G({\bf k} + {\bf q}, i\omega_{n} + i\omega_{m})M]\\&&-
    \nonumber \frac{4W_{0}^{2}}{\beta}
   \sum_{n,{\bf k}\in {\rm rbz}}\sin k_{x} \sin ({\bf k+q})_{x}\\&& {\rm Tr}
    [G({\bf k}, i\omega_{n})N
    G({\bf k} + {\bf q}, i\omega_{n} + i\omega_{m})N]
\label{eq:rhos}
\end{eqnarray}
In Eq.~\ref{eq:rhos}, the  $G$'s are the same matrix Green functions
used in the calculation of the spin-spin correlation function and 
$M$ and $N$ are the matrices
    \begin{eqnarray}
    \nonumber M &=&
  \left( \begin{array}{cccc} 1& 0 &
  0 & 0 \\
 0 & -1 & 0 & 0\\
  0 & 0 & 1 &0 \\
   0 &0 & 0 & -1 \\
    \end{array} \right), 
   \end{eqnarray}
   
   \begin{eqnarray}
\nonumber N &=&
  \left( \begin{array}{cccc} 0& 1 &
  0 & 0 \\
 -1 & 0 & 0 & 0\\
  0 & 0 & 0 &-1 \\
   0 &0 & 1 & 0 \\
    \end{array} \right) 
   \end{eqnarray}

The superfluid stiffness (and therefore the inverse
square penetration depth)
at $T=0$, evaluated in the entire doping range, is given by
  \begin{eqnarray}
  \nonumber\frac{n_s}{m^*}&=& -32t^{2}
\sum_{{\bf k}\in {\rm rbz}}\sin^{2}k_{x} W_{{\bf
k}}^{2}\left(1+\frac{\epsilon_{{\bf k}}^{2}+
   \Delta_{{\bf k}}^{2}+W_{{\bf k}}^{2} 
    -\mu^{2}}{E_{1}({\bf k})E_{2}({\bf k})}\right) \\ 
  \nonumber &\times&\frac{1}{[E_{1}({\bf k})+E_{2}({\bf k})]^{3}}
    (1+{\cal O}(W_{0}^2/t^2))\\ \nonumber
   &+&8t^{2} {\rm lim}_{\beta \rightarrow \infty}
\sum_{{\bf k}\in {\rm rbz}}\sin^{2}k_{x}
\frac{\epsilon_{{\bf k}}^{2}}{W_{{\bf k}}^{2}+\epsilon_{{\bf
k}}^{2}} \\ \nonumber
  &\times&  \left(\frac{df(E_{1}({\bf k}))}{dE_{1}({\bf k})}
+ \frac{df(E_{2}({\bf k}))}{dE_{2}({\bf k})}\right)
(1+{\cal O}(W_{0}^2/t^2))\\ \nonumber
   &-& 4t\sum_{{\bf k}\in {\rm rbz}}\cos k_{x}
\left(1+\frac{\epsilon_{{\bf k}}^{2}+
   \Delta_{{\bf k}}^{2}+W_{{\bf k}}^{2} 
    -\mu^{2}}{E_{1}({\bf k})E_{2}({\bf k})}\right)\nonumber\\ 
&\times& \frac{\epsilon_{{\bf k}}}{E_{1}({\bf k})+E_{2}({\bf k})}
(1+{\cal O}(W_{0}^2/t^2))
\label{eq:ds}  
  \end{eqnarray}
where $E_{1}({\bf k})$ and $E_{2}({\bf k})$ are
the two energy-values given before.
The first two terms in Eq.~\ref{eq:ds} come from
the paramagnetic current response, and the last
term is the kinetic energy (or the diamagnetic term).
At finite, but low temperatures, one can show that
\begin{eqnarray}
\frac{n_s}{m^*}(T)&-&\frac{n_s}{m^*}(0)\approx 8t^{2} 
\sum_{{\bf k}\in {\rm rbz}}\sin^{2}k_{x}
\frac{\epsilon_{{\bf k}}^{2}}{W_{{\bf k}}^{2}+\epsilon_{{\bf
k}}^{2}}\nonumber \\ 
   &\times&  \left(\frac{df(E_{1}({\bf k}))}{dE_{1}({\bf k})}
+ \frac{df(E_{2}({\bf k}))}{dE_{2}({\bf k})}\right)
(1+{\cal O}(W_{0}^2/t^2)).
\label{eq:ns_T}
\end{eqnarray}

The ${\cal O}(W_{0}^2/t^2)$ terms derive from
the modification of both the kinetic energy and
current operators in the low-energy effective (Hartree-Fock)
Hamiltonian associated with DDW order. The modified current operator
gives rise to the matrix
$N$ in (\ref{eq:rhos}). Since these terms
are small, ${\cal O}(W_{0}^2/t^2)$, we neglect them in
the following discussion.

Equation \ref{eq:ds} has several interesting properties.
First, for $\Delta_{{\bf k}}=\Delta_{0}\neq 0$ and $W_{{\bf k}}=W_{0}$,
that is, for competing CDW and $s$-wave superconductivity,
the terms involving the Fermi functions do not contribute
at zero-temperature. At $\mu=0$, $E_{1}({\bf k})=E_{2}({\bf k})$,
and the same coherence factor in the 
other two terms becomes two. Then, by a
partial integration one can show that (denoting the first
term by $K^{\rm para}$ and the third term by
$K^{\rm dia}$), $ K^{\rm para}=-\frac{W_{0}^{2}}{\Phi_{0}^{2}}K^{\rm dia}$,
and the full kernel is given by
$K^{\rm total}=K^{\rm para}+K^{\rm dia}=
K^{\rm dia}(1-\frac{W_{0}^{2}}{\Phi_{0}^{2}})$.
When the two orders are of
$s$-wave type \cite{Fradkin},
$K^{\rm total}$ yields a superfluid density which
is maximum
when $W_{0}$=0 ({\it i.e}, $\Delta_{0}=\Phi_{0}$) 
and zero when $W_{0}$=$\Phi_{0}$ ({\it i.e},
$\Delta_{0}$=0)\cite{Miller}.
      
Returning to $d$-wave order parameters,
at $\Delta_{0}=0$, Eq.~\ref{eq:ds} reduces to the superfluid density
of the DDW state, which is, of course, zero.
The second term, which contains the derivatives of the Fermi 
functions is crucial for the cancellation in this case.
For $\Delta_{0}$ {\it and} $\mu$ finite, the second term
does not contribute at zero temperature (except for
a contribution from a single point in 
k-space; but the finite and almost constant contribution
from this is ignored here for simplicity). The system
now acquires a finite superfluid density.

The superfluid density derived from Eq.~\ref{eq:ds}
can be cast as a function of a single variable, $x$, by 
expressing the three parameters $W_{0}$, $\Delta_{0}$ and
$\mu$ in terms of the doping fraction. 
For the dependence of  $\mu$ on
$x$, we use $\mu\simeq -x^{2}$ (in units in which $2t=1$)
as in reference\cite{Furukawa}, but other
reasonable dependences 
should give similar results ({\em cf.} the discussion in the previous section). 
Hence,
the results of this section are valid and relevant
for the cuprates even if $\mu$ is
not pinned close to zero in the underdoped regime.
For $W_{0}$ and $\Delta_{0}$ we choose the relations given in
 Eq.~\ref{eq:doping} and
Eq.~\ref{eq:W0} only because they have received some 
recent experimental support \cite{Loram2}. But it's worth-emphasizing
that the qualitative features that we extract from Eq.~\ref{eq:ds}
for the behavior of the superfluid density with $x$ are fairly 
independent of the precise functional forms of $W_{0}$ and $\Delta_{0}$.
The only input that is needed is the existence of DDW order,
with diminishing strength with $x$, and complementary development
of the DSC order. The DDW order eats away part of the superfluid density
from an otherwise pure DSC system even within the superconducting dome
in the $T-x$ phase diagram. With increasing
$x$, the DDW order weakens, hence
the superfluid density increases in a model-independent manner.

In Fig.~\ref{fig:2} we plot $n_s/m^*$ against 
the doping-fraction $x$ using Eq.~\ref{eq:doping} and
Eq.~\ref{eq:W0} for $W_{0}$ and $\Delta_{0}$. To illustrate
the qualitative robustness of our result, we also
show, by the dashed curve, the same plot for  $W_{0}$ dispersing quadratically 
with $x$ between
the same two end points as in Eq.~\ref{eq:doping}. 
 As can be seen from the figure the zero-temperature superfluid
density, in a 
model-independent manner, shows a rapid drop in
the underdoped regime, similar to that observed in experiments.  

 The leading temperature dependence of the superfluid stiffness
can be evaluated from Eq.~\ref{eq:ns_T}. At temperatures
 much smaller than the relevant energy scales $W_{0}$ and $\Delta_{0}$,
 only the nodal regions close to the points $(\pi/2, \pi/2)$ and
 symmetry-related points will contribute in the suppression
 of the superfluid weight. By expanding  around those points, one can see that
 the leading temperature dependence is indeed linear for the optimally
 and moderately doped samples in a fairly wide range of temperatures.
 For these doping concentrations, where $\Delta_{0}$ is larger than
 or comparable to $W_{0}$, $W_{0}$ plays a subleading role to $\Delta_{0}$
 in determining the temperature dependence of the suppression. This is
 due to the peculiar band structure of the problem.
 
\begin{figure}[bht] 
   \narrowtext 
   \begin{center} 
   \leavevmode 
%\vspace{.2cm} 
   \noindent 
   \hspace{0.3 in} 
   \centerline{\epsfxsize=2.7in \epsffile{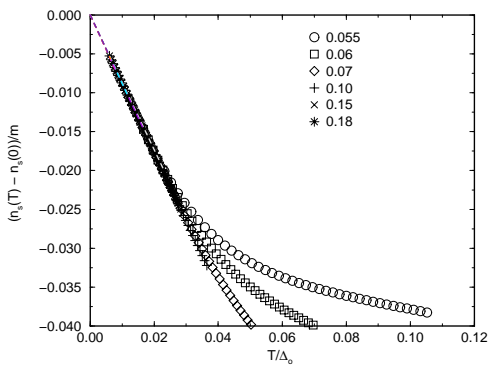}} 
%\epsfxsize=3.3in 
%\epsfysize=2.in 
%\epsfbox{} 
%\vspace{.2cm} 
   \end{center} 
   \caption 
{ The temperature-dependent part of the superfluid stiffness plotted
against scaled temperature for six values of the doping.
The doping concentrations are indicated in the legends. The values of
$W_{0}$ and $\Delta_{0}$ for a given doping are
derived from Eqs.~\ref{eq:doping} 
and \ref{eq:W0}. $\mu$ is derived from $\mu =-x^2$ and $t=0.5$ eV. The dotted li
ne is an
extrapolation of the results to zero temperature.} 
   \label{fig:4} 
   \end{figure}

 On the other hand, for the heavily underdoped samples the situation
 is quite different. Critical fluctuations are clearly important
 to determine any low-temperature property of the system, for it is close
 to a quantum critical point. Even if we ignore these fluctuations, and
 rely strictly on our mean-field results, the conclusions are very
 different from moderate or optimally doped samples. Though in the
 asymptotically low temperature regime, the depletion of the superfluid
 density is linear in temperature, there is an intermediate
 temperature range over which the
 suppression of the superfluid density actually behaves as
 $\sqrt{T}$. As the DDW gap is much larger than the superconducting
 gap in these heavily underdoped samples, $W_{0}$ crosses over to
 produce the leading contributions in the expansion of Eq.~\ref{eq:ns_T} around the
nodes, and is eventually responsible
 for the $\sqrt{T}$ behavior of the suppression. All of this is summarized
 in Fig.~\ref{fig:4}, where we have plotted the temperature-dependent
 part of the superfluid stiffness with temperature scaled by the
 superconducting gap, for six values of the doping fraction $x$. The
 plot shows $\sqrt{T}$ behavior for the heavily underdoped systems
 for the experimentally relevant temperatures\cite{Panago}. For asymptotically
 low temperatures for these doping values, and for a fairly wide range of
 temperatures for moderate or optimal doping, the data
 show exact collapse on a single straight line signifying a unique
 low-temperaure slope.

\section{Discussion}

The cuprate phase diagram proposed
in reference\cite{Sudip} is reproduced here in Fig.~\ref{fig:3} for
easy reference; we only show the relevant portion of the phase diagram,
ignoring the complex set of competing charge ordered states in the
underdoped regime as well as the spin glass phase.

{}From our calculation, the behavior of the $T=0$
superfluid density with $x$ is as 
expected from experiments in the underdoped
regime of the high-$T_{c}$ cuprates.
The striking rapid suppression of the superfluid
density (and sharp increase of the London
penetration depth) below $x\simeq 0.2$ is naturally
explained by the emergence of the singlet
DDW at that doping-fraction. The behavior of the
superfluid density near $T_c$ or near $x\approx 0.05$ (the
lowest doping at which superconductivity occurs)
will be strongly affected by thermal or quantum phase fluctuations,
so our Hartree-Fock results will be suspect in those regimes.
However, away from these critical points, we might
expect our calculation to be on solid footing, and it
is encouraging to see that it agrees with
experimental measurements of $n_s$. The temperature 
dependence also captures the striking linearity shown in Fig.~\ref{fig:4}, and
the doping independence of the slope, along
 with 
a $\sqrt{T}$ dependence in an intermediate temperature regime in the heavily
underdoped  regime akin to the experiments\cite{Panago}.

The neutron scattering peaks observed in the
pseudogap and superconducting regimes are
broadly consistent with our calculations.
There are peaks at $(\pi,\pi)$
at energies which are controlled by
the single-particle gap and the doping.
There are no observable  peaks at $(\pi,0)$
and symmetry-related points because, from our analysis,
they would be at high energies
controlled by the band dispersion where
they are likely to be strongly damped.
This resolves an earlier puzzle of the analysis of
the staggered flux phase\cite{Affleck}. 
Though we have not presented here results for finite tempeartures,
from the phase diagram 
and our zero temperature theory, the
qualitative aspects at finite temperatures can
still be explained. However, the 
details of the peak
position in energy and its doping dependence is
beyond a simple Hartree-Fock calculation.

If one approaches along the path labeled (1)
in Fig.~\ref{fig:3},  at first one will 
find no structure in the spin-spin correlator,
typical of a Fermi liquid. Below
$T^{*}$, the correlator peaks at an energy $2W_{0}$, twice the maximum of the
DDW gap at the wavevector 
$\bf Q$. This is consistent with experiments.  Below $T_{c}$,
the intensity in the spin-spin correlator is amplified
as the  DSC
order develops, as discussed in Ref.~\cite{CK}.
Due to the coexistence of $DDW$ and $DSC$, the peak will be at a higher
energy --- $2\sqrt{ W_{0}^{2} + 
\Delta_{0}^{2}}$, shifted by an amount that depends
on the doping. To see this, recall that, for coexisting DDW and DSC,
the peak-energy shifts to higher values with $|\mu|$,
which increases with doping.

Consider now the path (2) in Fig.~\ref{fig:3}. 
In the DDW state, and for small chemical potential,
the neutron scattering intensity  
should exhibit a peak at wavevector $\bf Q$
and energy $2W_{0}$; recall that, for
DDW order, the peak-energy shows no movement with $|\mu|$, but
is destroyed by a large $|\mu| > W_{0}$. Since
the DDW gap varies with $x$ as in
Eq.~\ref{eq:doping}, the peak would shift to higher
energies as $x$ is decreased, 
if $\chi_{0}$
alone were responsible for the spin-fluctuation
spectrum. However,
the reduced Coulomb interaction
$\overline U$ is important since it shifts the
peaks to lower energies, broadening them simultaneously. For smaller $x$, due
to the lower density of the holes, the
effective $\overline U$ increases,
and, as a result,
as shown in Fig.~\ref{fig:1}, the peak
moves to lower energies.
Consequently, the peak-energies in this part of the 
phase diagram would be influenced by these two competing 
effects.

Consider the
path (3) in Fig.~\ref{fig:3}. In the coexisting region,
DDW plus DSC, the peak energy is $2\sqrt {W_{0}^{2} + 
\Delta_{0}^{2}}$, shifted by $|\mu|$, but now, in contrast to $T>T_{c}$, 
the magnitude of the total order parameter
remains constant as per Eq.~\ref{eq:W0}. Of course,
$\overline U$ must be important as well, modifying this conclusion.

\acknowledgments

We would like to thank Qiang-Hua Wang for pointing
out a minor error in an earlier version of this paper.
S. C. and S. T. are supported by NSF-DMR-9971138, and C. N. by 
NSF-DMR-9983544
and the A.P. Sloan Foundation.
The work of H. -Y. K. was conducted under
the auspices of the DOE, supported by funds
provided by the University of California for the conduct
of discretionary research by Los Alamos National Laboratory.

\end{document}